\documentclass[aps,prd,floatfixy,superscriptaddress,showpacs,showkeys]{revtex4}
\usepackage{amssymb}
\usepackage{amsmath}
\usepackage{amsfonts}
\usepackage{epsfig}
\usepackage{graphicx}
\usepackage{tabularx}
\usepackage{latexsym}
\usepackage{subfigure}
\usepackage{verbatim}
\usepackage{color}
\usepackage{braket}
\newcommand{\seq}{\begin{subequations}}
\newcommand{\sen}{\end{subequations}}
\newcommand{\eq}{\begin{eqnarray}}
\newcommand{\en}{\end{eqnarray}}

\def\ds{D^{\ast  0}}

\def\L2{\Lambda^2}

\def\eq{\begin{eqnarray}}
\def\en{\end{eqnarray}}

\usepackage[normalem]{ulem}

\renewcommand\sout{\bgroup \color{red} \ULdepth=-.5ex \ULset}

\def\ds{D^{\ast  0}}

\def\L2{\Lambda^2}

\def\eq{\begin{eqnarray}}
\def\en{\end{eqnarray}}

\def\L{{\cal L}}

\def\ds{d^*}
\def\vk{\vec{k}}
\def\vr{\vec{\rho}}
\def\vl{\vec{\lambda}}

\begin{document}

\title{\boldmath A study of $d^*(2380)\to d \pi\pi$ decay width  }

\author{Yubing Dong}
\affiliation{Institute of High Energy Physics, Chinese Academy of
Sciences, Beijing 100049, China} \affiliation{Theoretical Physics
Center for Science Facilities (TPCSF), CAS, Beijing 100049, China}

\author{Pengnian Shen}
\affiliation{College of Physics and Technology, Guangxi Normal
University, Guilin  541004, China} \affiliation{Institute of High
Energy Physics, Chinese Academy of Sciences, Beijing 100049, China}
\affiliation{Theoretical Physics Center for Science Facilities
(TPCSF), CAS, Beijing 100049, China}

\author{Fei Huang}
\affiliation{School of Physical Sciences, University of Chinese
Academy of Sciences, Beijing 101408, China}

\author{Zongye Zhang}
\affiliation{Institute of High Energy Physics, Chinese Academy of
Sciences, Beijing 100049, China} \affiliation{Theoretical Physics
Center for Science Facilities (TPCSF), CAS, Beijing 100049, China}

\date{\today}

\begin{abstract}
The decay widths of the $\ds\to d \pi^0\pi^0$ and $\ds\to d
\pi^+\pi^-$ processes are explicitly calculated in terms of our
chiral quark model. By using the experimental ratios of cross
sections between various decay channels, the partial widths of the
$\ds\to pn \pi^0\pi^0$, $\ds\to pn \pi^+\pi^-$, $\ds\to pp
\pi^0\pi^-$, and $\ds\to nn \pi^+\pi^0$ channels are also extracted.
Further including the
estimated partial width for the $\ds\to pn $ process, the total
width of the $\ds$ resonance is obtained. In the first step of the
practical calculation, the effect of the dynamical structure on the
width of $\ds$ is studied in the single $\Delta\Delta$ channel
approximation. It is found that the width is reduced by few tens of
MeV, in comparison with the one obtained by considering the effect
of the kinematics only. This presents the importance of such effect
from the dynamical structure. However, the obtained width with the
single $\Delta\Delta$ channel wave function is still too large to
explain the data. It implies that the $\ds$ resonance will not
consist of the $\Delta\Delta$ structure only, and instead there
should be enough room for other structure such as the hidden-color
(CC) component. Thus, in the second step, the width of $\ds$ is
further evaluated by using a wave function obtained in the coupled
$\Delta\Delta$ and CC channel calculation in the framework of the
Resonating Group Method (RGM). It is shown that the resultant total
width for $\ds$ is about 69 MeV, which is compatible with the
experimental observation of about 75 MeV and justifies our assertion
that the $\ds$ resonance is a hexaquark-dominated exotic state.
\end{abstract}



\maketitle

\section{Introduction}

In recent years, the CELSIUS/WASA and WASA@COSY Collaborations
successively reported the observation of a resonance-like structure
in the double pionic fusion channels $pn\to d\pi^0\pi^0$ and $pn\to
d\pi^+\pi^-$ when they studied the ABC effect and in the polarized
neutron-proton scattering~\cite {CELSIUS-WASA,CELSIUS-WASA1,ABC}.
They mentioned that because the width of the structure is rather
narrow, which is three more times smaller than $2\Gamma_{\Delta}$ in
the conventional $\Delta\Delta$ process, the observed data cannot be
explained by the contribution from either the Roper excitation or
the t-channel $\Delta\Delta$ process. Therefore, they
proposed a $d^*$ hypothesis, in which its quantum number, mass and
width are $I(J^P)=0(3^+)$, $M \approx 2.36$ GeV and $\Gamma \approx
80$ MeV~\cite{CELSIUS-WASA} (in their recent paper~\cite{Bashkanov},
they take averaged values over the results from elastic scattering
and two-pion production, $i.e.$ $M \approx 2.375$ GeV and $\Gamma
\approx 75$ MeV), respectively, to accommodate the data. Because
``the structure, containing six valence quarks, constitutes a
dibaryon, and could be either an exotic compact particle or a
hadronic molecule"~\cite{CERN}, this result causes physicists'
special attention.

In fact, the existence of the non-trivial six-quark configuration
with $I(J^P)=0(3^+)$ (called $d^*$ lately) has intensively been
studied since Dyson's estimation~\cite{Dyson}. A variety of methods
or models, such as group theory~\cite{Dyson}, bag quark
model~\cite{Thomas}, quark potential model~\cite{Oka,Wang,Yuan},
etc., have been employed to investigate the structure of $\ds$,
among which even some investigations produced a mass close to the
recent data, they are either not a dynamical calculation, or a
calculation without the width prediction or with an incorrect width
prediction. It should specially be noted that in one of those
papers~\cite{Yuan}, one performed a coupled channel dynamical
calculation in 1999 where a $\Delta\Delta$ channel and a
hidden-color channel (denoted by CC) are included and the predicted
mass is about $40-80$ MeV. This means that in this structure, there
might exist a six-quark configuration, which coincides with COSY's
assertion. Nevertheless, in that paper, the width of the state has
not been calculated.

After COSY reported their finding, many investigations have been
devoted to this aspect. There are mainly three kinds of models on
the structure of the $d^*$ resonance: a) It is a $\Delta\Delta$
resonance~\cite{Wang1}. The authors in Ref.~\cite{Wang1} performed a
multi-channel scattering calculation and obtained a binding energy
about 71 MeV with respect to the $\Delta\Delta$ threshold and a
width about 150 MeV where $\Gamma_{NN}=14$ MeV and
$\Gamma_{inel}=136$ MeV. b) It is dominated by a ``hidden-color"
six-quark configuration. Bashkanov, Brodsky and
Clement \cite{Brodsky} argued in 2013 that this hidden-color structure
is necessary for understanding the strong coupling of $d^*$ to $\Delta\Delta$. Later,
Huang and his collaborators made an explicit dynamic calculation in
the framework of the Resonating Group Method (RGM)~\cite{Huang} and
showed a binding energy of about $84$ MeV and almost $67\%$ of
``hidden-color" configuration in $d^*$. This implies that $\ds$ is
probably a 6-quark dominated exotic state. c) It is a result of the
$\Delta N \pi$ three body interaction~\cite{Gal}. In order to
justify which one of these three is more reasonable, a detailed
calculation, especially the decay width, should be performed and further
experimental investigation should be carried out.

In this paper, we focus on $d^*$ width study. We would firstly exam
the effect of the dynamical structures of the $\ds$
and deuteron bound states on the decay width of $\ds$, and
consequently fetch out the contribution from the $\Delta\Delta$
structure of $\ds$ with $J^P=3^+$. Then, we would estimate the total
width of $\ds$ by including the contributions from other possible
decay channels. At the beginning, we temporarily
assume that $d^*$ is composed of the $\Delta\Delta$ structure only.
In the calculation, the extended chiral SU(3) quark model is
employed, because this constituent quark potential model can
successfully reproduce the spectra of baryon ground states, the
binding energy of the deuteron, the nucleon-nucleon (NN),
Kaon-nucleon (KN) scattering phase shifts, and the hyperon-nucleon
(YN) cross sections (for details see
Refs.~\cite{Zhang1,Zhang2,Huang2}). With the same set of model
parameters fixed in explaining the above mentioned data, the bound
state problem of the $\Delta\Delta$ system is solved and the
realistic wave functions of $\ds$ and deuteron are obtained via the
dynamical RGM calculation. With these wave functions, the two-pion
decay width of $\ds(2380)$ in the process of $\ds\to d \pi\pi$ is
calculated on the quark level, where the chiral effective Lagrangian
of quark-quark-pion is employed. In terms of the experimental data
of other observed decay channels such as $d^* \to np\pi^0\pi^0$,
$d^* \to np\pi^+\pi^-$, $d^* \to nn\pi^+\pi^0$, $d^* \to pp\pi^0\pi^-$, etc.,
the total width of $\ds$ is estimated, and
the role of dynamical structures to the decay width is analyzed. The
result with the single $\Delta\Delta$ channel assumption exhibits
the importance of the dynamical structure effect which reduces the
decay width by about few tens of MeV. However, the width is still
larger than the experimentally observed value, so that the other
structure in $\ds$ should further be considered. Subsequently, we
evaluate the width of $\ds$ with the wave function obtained in the
coupled $\Delta\Delta$ and CC channel RGM calculation~\cite{Huang}.
The resultant total width of $\ds$ is about $69$ MeV, which is
compatible with the experimental data. In the next section, the
formulism is briefly given. The numerical results and discussion are
presented in the final section.

\section{Brief formulism}

Referring to Ref.~\cite{Buchmann}, the phenomenological effective
Hamiltonian for the quark-quark-pion interaction in the
non-relativistic approximation is \eq {\cal
H}_{qq\pi}=g_{qq\pi}\vec{\sigma}\cdot \vk_{\pi}\tau\cdot\phi\times
\frac{1}{(2\pi)^{3/2}\sqrt{2\omega_{\pi}}}, \en where $g_{qq\pi}$ is
the coupling constant, $\phi$ stands for the $\pi$ meson field,
$\omega_{\pi}$ and $\vk_{\pi}$ are the energy and three-momentum of
the $\pi$ meson, respectively, and ${\bf \sigma} ({\bf \tau})$
represents the spin (isospin) operator of a single quark. The wave
functions are \eq \mid N> \,=\frac{1}{\sqrt{2}}\Big
[\chi_{\rho}\psi_{\rho}+\chi_{\lambda} \psi_{\lambda}\Big
]\Phi_N(\vr,\vl) \en for the nucleon and \eq \mid
\Delta>\,=\chi_s\psi_s\Phi_{\Delta}(\vr,\vl) \en for the
$\Delta(1232)$ resonance. In Eqs. (2-3), $\chi$ and $\psi$ stand for
their spin and isospin wave functions, $\Phi_N(\vr,\vl)$ and
$\Phi_{\Delta}$ are the spatial wave functions of the nucleon and
$\Delta$ resonance, respectively, and $\rho$ and $\lambda$ are the
Jacobi coordinates for the internal motion. Then, the decay width
for $\Delta\to \pi N$ reads \eq \Gamma_{\Delta\to\pi N}
=\frac{4}{3\pi}k_{\pi}^3(g_{qq\pi}I_o)^2\frac{\omega_N}{M_{\Delta}},
\en where $\omega_{\pi,N}=\sqrt{M_{\pi,N}^2+\vk_{\pi}^2}$ are the
energies of the pion and nucleon, respectively, $k_{\pi}\sim 0.229$
GeV, and $I_o$ denotes the spatial overlap integral of the internal
wave functions of the nucleon and the $\Delta$ resonance. By fitting
the measured width of 117~MeV for $\Delta_{3/2^+}(1232)$~\cite{PDG},
one gets $G=g_{qq\pi}I_o\sim 5.41$ GeV$^{-1}$ which is the product
of the coupling constant $g_{qq\pi}$ and the spatial integral $I_0$.

Now using the knowledge of ${\cal M}_{\Delta\to\pi N}$ obtained
above, we can estimate the decay width in the $\ds\to d \pi\pi $
process. The transition matrix element between the initial state
$\ds$ and the final state $d\pi^0\pi^0$ can be written as
\begin{eqnarray}
{\cal M}^{\pi^0\pi^0}_{if}&=&\frac{1}{\sqrt{3}}\sum
F_1F_2k_{1,\mu}k_{2,\nu}I^0_SI^0_I C_{1\nu,1\mu}^{jm_j}C_{3m_{d^*},
jm_j}^{1m_{d}}
 \nonumber \\
&&\times\int ~d^3q~\Big [
 \frac{\chi^*_d(\vec{q}-\frac12\vec{k}_{12})}{E_{\Delta}(q)- E_N(q-k_1)-\omega_1}
+\frac{\chi^*_d(\vec{q}+\frac12\vec{k}_{12})}{E_{\Delta}(q)- E_N(q-k_2)-\omega_2}
 \nonumber \\
&&~~~~~+\frac{\chi^*_d(\vec{q}+\frac12\vec{k}_{12})}{E_{\Delta}(-q)-E_N(-q-k_1)-\omega_1}
+\frac{\chi^*_d(\vec{q}-\frac12\vec{k}_{12})}{E_{\Delta}(-q)-E_N(-q-k_2)-\omega_2}
\Big ] \chi_{d^*}(\vec{q}), \label{matrixa}
\end{eqnarray}
where $i$ and $f$ stand for the initial $d^*$ state with quantum
numbers $((S m_S) = (3 m_{d^*}))$ and the final deuteron state with
$((S m_S) = (1 m_d))$, respectively, $I_{S(I)}^0$ is the spin
(isospin) factor shown in the appendix,
$F_{1,2}=F(k^2_{1,2})=\frac{4G}{(2\pi)^{3/2}\sqrt{\omega_{1,2}}}$,
$\vec{k}_{12}=\vec{k}_1-\vec{k}_2$,
$\omega_{1,2}=\sqrt{m_{\pi}^2+\vec{k}_{1,2}^2}$. $\chi_d(\vec{q})$
and $\chi_{d^*}(\vec{q})$ are, respectively, the relative wave
functions of the final deuteron (between the two nucleons) and the
initial $d^*$ (between the two-$\Delta$s) where
$\vec{q}=\frac12(\vec{p}_1+\vec{p}_2+\vec{p}_3-\vec{p}_4-
\vec{p}_5-\vec{p}_6)$ with $\vec{p}_i$ being the momentum of the
$i$-th quark.
Four terms in the bracket of
Eq.~(\ref{matrixa}) are related to the propagators of four
sub-diagrams in Fig. 1. \vspace{0.5cm}

\begin{figure}[htbp]
\centering \includegraphics [width=7cm, height=8cm]{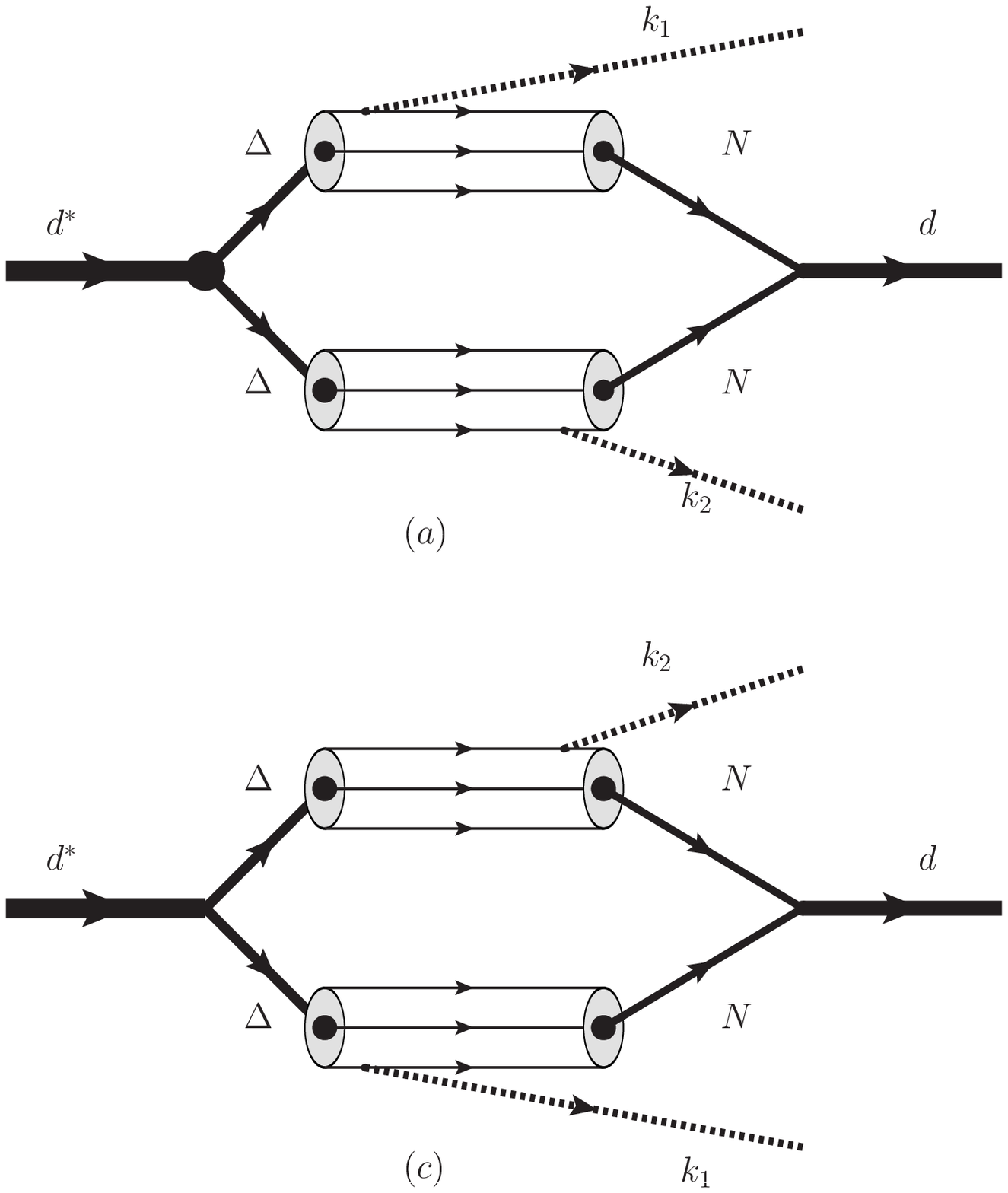}
{\hskip 1cm}
\includegraphics [width=7cm, height=8cm]{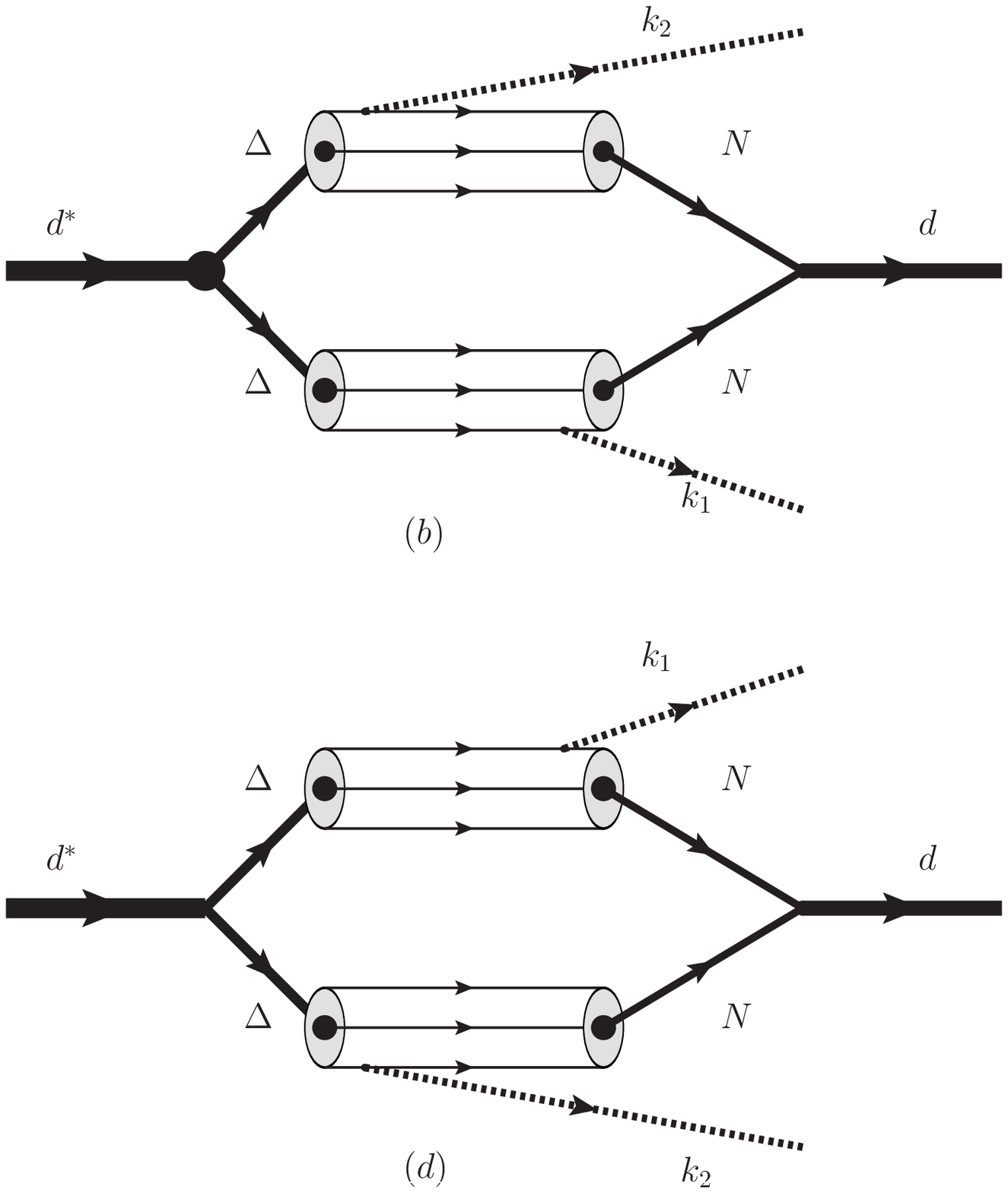}
\caption{Four possible emission ways in the decay of the $\ds$
resonance composed of the $\Delta\Delta$ structure only. Two pions
with momenta of $\vec{k}_{1,2}$ are emitted from one of the three
quarks in 2 $\Delta$s, respectively.}
\end{figure}

With the transition matrix element ${\cal M}_{if}^{\pi^0 \pi^0}$,
the decay width of $\ds$ in the $\ds \to d \pi\pi$ channel can be
evaluated by \eq \Gamma_{d^*\to d\pi^0\pi^0} =\frac{1}{2!}\int
d^3k_1d^3k_2d^3p_{d}(2\pi) \delta^3(\vec{k}_1+\vec{k}_2+\vec{p}_d)
\delta(\omega_{k_1}+\omega_{k_2}+E_{p_{_d}}-M_{\ds})\mid
\overline{{\cal M}}^{\pi^0\pi^0}_{if}\mid ^2, \label{gamma} \en
where $\omega_{k_1,k_2}$ are the energies of the two outgoing pions,
$E_{p_{_d}}$ is the energy of the outgoing deuteron with momentum
$p_{d}$, and the bar on the top of the transition
matrix ${\cal M}_{if}^{\pi^0 \pi^0}$ means
that this matrix element is averaged over the initial states and the
summed over the final states. The factor of $2!$ is due to the two
identical pions in the final states.

\begin{figure}[htbp]
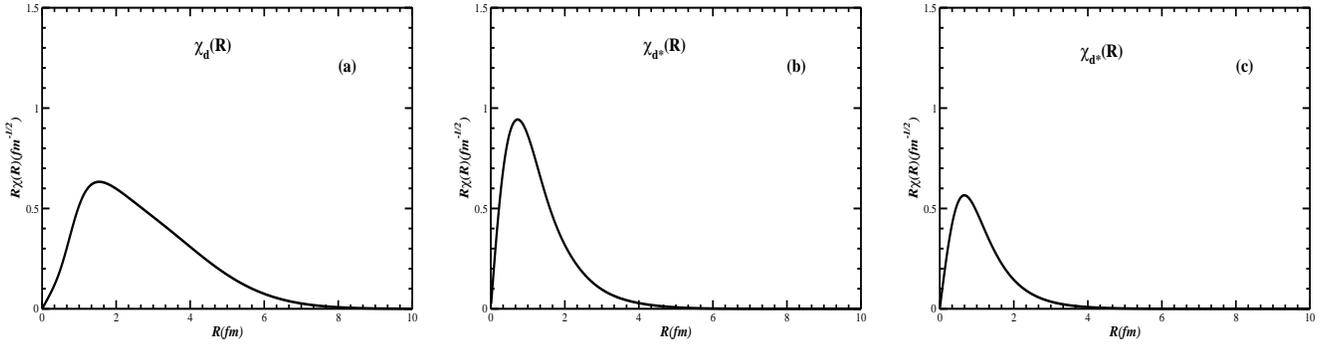

\centering
\includegraphics [width=5.5cm, height=4.5cm]{deuwave.eps}
{\hskip 0.25cm}
\includegraphics [width=5.5cm, height=4.5cm]{dswave.eps}
{\hskip 0.25cm}
\includegraphics [width=5.5cm, height=4.5cm]{dswave2.eps}
\caption{Relatively wave functions in $S-$wave in the extended
chiral SU(3) quark model: (a) for deuteron, (b) for
$\chi_{\Delta\Delta}$ in the single $\Delta\Delta$ channel case for
$d^*(2380)$, (c) for $\chi_{\Delta\Delta}$ in the coupled
$\Delta\Delta$ and CC channel case for $d^*(2380)$. } \label{wf}
\end{figure}

In
the practical decay width calculation, one needs the explicit
deuteron and $\ds(2380)$ relative wave functions. These wave
functions can usually be taken from the realistic solutions of the
system considered. In this work, we obtain them by dynamically
solving the RGM equation in the extended chiral $SU(3)$ quark
model~\cite{Huang} where the binding energy of deuteron is $\epsilon
= 2.2$ MeV and the binding energy of $\ds$ is $\epsilon\approx 62$
MeV in the single $\Delta\Delta$ channel approximation and
$\epsilon\approx 84$ MeV if the CC channel is further considered,
and consequently, the mass of $\ds$ is
$M_{\ds}=2M_{\Delta}-\epsilon$.
In the coordinate space, the wave
functions of the deuteron and $\ds$ systems can also be expressed,
respectively, as \eq
\Psi_d ~&=& [~\phi_N(\xi_1,\xi_2)~\phi_N(\xi_4,\xi_5)~\chi_d(R)~]~\zeta_{(SI)=(10)},\\
\nonumber
\Psi_{d^*} &=&
[~\phi_{\Delta}(\xi_1,\xi_2)~\phi_{\Delta}(\xi_4,\xi_5)~
\chi_{\Delta\Delta}(R)~ +~
\phi_{C}(\xi_1,\xi_2)~\phi_{C}(\xi_4,\xi_5)~
\chi_{CC}(R)~]~\zeta_{(SI)=(30)}, \en where
$\phi_N,~\phi_{\Delta},~\phi_{C}$ denote the internal wave functions
of $N,~\Delta,~C$ (color-octet particle) in the coordinate space,
$\chi_d$ describes the relative wave function of the deuteron,
$\chi_{\Delta\Delta}$ and $\chi_{CC}$ represent the relative wave
functions between $\Delta$s and $C$s (in the single $\Delta\Delta$
channel case, the CC component is absent), and $\zeta_{(SI)}$ stands
for the spin-isospin wave function of the corresponding system.
It should be specially mentioned that in the form of such a wave
function, normally called channel wave function~\cite{Huang}, the
totally anti-symmetric effect is implicitly included in the
resultant relative wave function by solving the RGM equation and
then projecting to the physical states. The
channel wave functions of relevant systems are plotted in
Fig.~\ref{wf}.

\vspace{1cm}

For the sake of convenience, we expand the relative wave function in
the following: \eq \chi(R)=\sum^4_{i\,=\,1}~c_i\,
\exp\left(-\frac{R^2}{2b_i^2}~\right). \en We would also mention
that the $D$-wave contribution is omitted due to its relevant
smaller contribution, although both the $S$- and $D$-wave functions
exist in our resultant wave functions.

With these wave functions, two additional assumptions are employed
in the estimation of the decay width in $\ds\rightarrow d\pi\pi$.
One is associated to the energy denominators in Eq.~(\ref{matrixa}),
where the pole position is simply taken, as usually did in K-matrix
approximation approach. The other one is related to the directions
of the two outgoing pions. Since the experimental data show that the
angle between two outgoing pions is almost zero, namely, the pions
are propagating in the same direction, we can employ a bilinear
condition for the momenta of these pions, $\vk_1\parallel
\vk_2\parallel {\hat z}$, so that the calculation can be much more
simplified without loosing the major characters of such a process.

\section{Numerical results and discussion}

In the calculation, the masses of deuteron, $\Delta$, nucleon and
pion are taken from Particle Data Group~\cite{PDG}. The mass of
$d^*$ is $M_{d^*}=2M_{\Delta}-\epsilon$ with $\epsilon$ being
$80-90$ MeV for the single $\Delta\Delta$ channel case and $\sim 84$
MeV for the coupled $\Delta\Delta$ and CC channel case,
respectively. The value of $G$ is already fixed by using the
$\Delta\to \pi N$ decay data. The decay width in the $\ds\to d
\pi\pi $ process with realistic wave functions from the RGM
calculation can numerically be obtained by using Eq.~(\ref{gamma}).

Moreover, the experimental data~\cite{CELSIUS-WASA,CELSIUS-WASA1,WASA1,WASA2}
and one of theoretical calculations~\cite{Oset} showed that for the $\ds$
resonance at $\sqrt{s}=2.37$ GeV, the decay cross section in the
$\ds\to pn\pi^+\pi^-$ process is about $0.20$ mb which is comparable
with that of $0.24$ mb in the $\ds\to d\pi^0\pi^0$ process, and the
decay cross section in the $\ds \to pp\pi^0\pi^-$ process
(also its mirrored channel $\ds \to nn \pi^+ \pi^0$)
has a visible value of about $0.10$ mb as well. Therefore,
contributions in these processes should also be accounted for in the
$\ds$ width estimation. Using Breit-Wigner formulism and those cross
section data, one estimated the branching ratios of various decay
modes~\cite{WASA3,Bashkanov}. For reference, we tabulate them in the
second last column in Table~\ref{tab:width}. In order
to consider the effect of isospin symmetry breaking of pions,
namely the mass difference between $\pi^{\pm}$ and $\pi^0$,
we calculate the cross section in the $\ds\to
d\pi^+\pi^-$ process explicitly. The obtained cross section in this
calculation is about 1.83 times that in the $\ds\to d\pi^0\pi^0$
process, which is slightly larger than the multiple of 1.6 estimated
in Ref.~\cite{WASA3,Bashkanov}. Based on resultant
decay widths for $\ds\to d\pi^0\pi^0$ and $\ds\to d\pi^+\pi^-$ and
cross sections mentioned above, we get the branching ratios, and
consequently the partial decay widths, for all possible decay modes.
Finally, we achieve the total width of $\ds$.

In order to see the effect of the dynamical structure on the decay
width, we calculate the width in the single $\Delta\Delta$ channel
case with $\epsilon=80$ or $90$ MeV (the corresponding mass of $\ds$
is about $2384$ MeV or $2374$ MeV, respectively) to compare with the
result reported by Bashkanov et al.~\cite{Brodsky}, where the decay
width is reduced by the phase space effect only. We tabulate
resultant decay widths for all the possible channels
in Table~\ref{tab:width}.
\begin{table}[htbp]
\begin{center}
\caption{ Decay width}
\begin{tabular}{|c||c|c|c|c||c|c|}\hline
&\multicolumn{4}{|c||}{$Ours$} &\multicolumn{2}{|c|}{$Expt.$}\\
\cline{2-5}
&\multicolumn{2}{|c|}{$~~\Delta\Delta~~$}
&\multicolumn{2}{|c||}{~~two channels $\Delta\Delta+{\rm CC}~~$}
&\multicolumn{2}{|c|}{\cite{WASA1,WASA2,WASA3,Bashkanov}}\\
\hline
$~~M_{d^*}({\rm MeV})~~$ &$~~~~2384~~~~$ &$~~~~2374~~~~$
&\multicolumn{2}{|c||}{$~~2380~~$}
&\multicolumn{2}{|c|}{$~~2375~~$}\\
\hline\hline
channel & {$~~~\Gamma({\rm MeV})~~~$} &{$~~~\Gamma({\rm MeV})~~~$}
&$~~~~~~~{\cal B}_r~~~~~~~$
&$~~~\Gamma({\rm MeV})~~~$  &$~~~~~~~{\cal B}_r~~~~~~~$ &$~~~\Gamma({\rm MeV})~~~$\\
\hline\hline
$~~d^*\to d\pi^0\pi^0~~$ &$22.6$ &$17.0$ &$13.3\%$ &$9.2$  &$14(1)\%$ &$10.2$\\
\hline
$~~d^*\to d\pi^+\pi^-~~$ &$41.5$ &$30.8$ &$24.3\%$ &$16.8$  &$23(2)\%$ &$16.7$\\
\hline
$~~d^*\to pn\pi^0\pi^0~~$ &$18.8$ &$14.2$ &$11.3\%$ &$7.8$  &$12(2)\%$ &$8.7$\\
\hline
$~~d^*\to pn\pi^+\pi^-~~$ &$47.1$ &$35.4$ &$27.8\%$ &$19.2$  &$30(4)\%$ &$21.8$\\
\hline
$~~d^*\to pp \pi^0\pi^-~~$ &$9.4$ &$7.1$ &$5.65\%$ &$3.9$   &$6(1)\%$ &$4.4$\\
\hline
$~~d^*\to nn \pi^+\pi^0~~$ &$9.4$ &$7.1$ &$5.65\%$ &$3.9$   &$6(1)\%$ &$4.4$\\
\hline
$~~d^*\to pn~~$ &$18.8$ &$14.2$ &$12.0\%$ &$8.3$  &$12(3)\%$ &$8.7$\\
\hline
$~~Total~~$ &$167.6$ &$125.8$ &$99.9\%$ &$69.1$  &$103(14)\%$ &$74.9$\\
\hline
\end{tabular}
 \label{tab:width}
\end{center}
\end{table}
From this table, one sees that in the single $\Delta\Delta$ channel
calculation, no matter in which case ($M_{\ds}=2384$~MeV or
$M_{\ds}=2374$~MeV), the resultant total width of $\ds$ justifies
the fact that in a composite system, due to the binding behavior,
namely the attractive interaction between constituents, the decay
width of the system is much smaller than the total decay widths of
its constituents if they were assumed to be free particles. And even
more, deeper binding would cause narrower width. This feature is
reasonable, because the width is not only related to the phase
space, but also depends on the overlap of the wave functions of the
bound states $d^*$ and deuteron. In comparison with the estimated
width of about $160$ MeV with the binding energy of $90$ MeV by
Bashkanov et al.~\cite{Brodsky}, where the effect of the phase space
is considered only, the contribution to the width from the dynamical
structure of the system is about few tens of MeV. This tells us how
important the effect of the dynamics on the width of an unstable
composite system is, namely, the decay width is not only related to
the phase space, but also depended on the dynamical structure of the
system. It also shows that the width of $\ds$ in the single
$\Delta\Delta$ channel case where the mass of $\ds$ coincides the
experimental data of $2384$ MeV still far exceeds the experimental
value of $75$ MeV. This means that the $\Delta\Delta$ structure
alone cannot provide a reasonable width of $\ds$.

With the same scenario, we further exam the width contributed by the
$\Delta\Delta$ component in $d^*$ if $d^*$ has the
$\Delta\Delta+${\rm CC} structure proposed in
Refs.~\cite{Yuan,Huang}. The results are also tabulated in
Table~\ref{tab:width}. It shows that with the wave function of the
$\Delta\Delta$ component in Ref.~\cite{Huang}, the decay widths for
the $\ds \to d \pi^0\pi^0 $ and $\ds \to d \pi^+\pi^- $ modes are
about $9$ MeV and $17$ MeV, respectively. If we further consider the
$\ds \to pn \pi\pi$, $pp\pi^0\pi^-$, $nn\pi^+\pi^0$, and $NN$ modes,
the total width would be about $69.1$ MeV.

Here we would like to mention that in the RGM calculation, the trial
wave function of the $\ds$ system is assumed to have two major
components, $\Delta\Delta$ and CC, which are totally
anti-symmetrized. Solving the RGM equation, one obtains the relative
wave functions of the system. By projecting the resultant wave
function onto the cluster internal wave function in each component,
we get the inter-cluster relative wave function, namely the channel
wave function, for corresponding channel. Now, the contribution
from the CC channel via the quark exchange is included in the
projected wave function (or channel wave function)
$\chi_{\Delta\Delta}(R)$ already~\cite{Huang}. We should specially
emphasis that the channel wave functions obtained in Eq. (7)
are orthogonal to each other. Therefore, in the
lowest order, by using this channel wave function, there is no quark
exchange between the two physical particles and thus the colored
clusters (color octet) cannot turn into the un-colored clusters
(color singlet). As a consequence, the width contributed by the
projected CC component would almost be zero. Combining this point
with the contribution from the $\Delta\Delta$ component, one sees
that total width of $\ds$ in our $\Delta\Delta$ +CC model is about
$69.1$ MeV, which is compatible with the experimental data of $75$
MeV. Apparently, because the fraction of the wave function of the CC
component in our $\Delta\Delta $+ CC model is about $67\%$, the
resultant width of $\ds$ justifies our assertion that the $\ds$
resonance is a hexaquark-dominated exotic state.

%Finally, it should also be mentioned that the existence of $\ds$
%should further be checked in other experimental processes. Now,
%except the $\gamma d$ reaction and $pp$ collision, BEPCII/BESIII
%Collaboration has accumulated a large amount of $\psi^{\prime}$
%events with its energy larger than $4.26$ GeV and has already
%detected the deuteron in the $\psi^{\prime}$ decay. Therefore, it is
%possible to measure $\ds$ in the $e^+e^-\to \psi^{\prime}( >
%4.26~{\rm GeV} )\to \ds+\bar{p}+\bar{n}\to d +\pi+ \pi
%+\bar{p}+\bar{n}$ processes where the quantum number of
%$\psi^{\prime}$ could be $I(J^P)=0(2^-,3^-,4^-)$~\cite{LIHB}. If one
%could observe $\ds$ in the data set accumulated by BEPCII/BESIII, it
%would definitely help us to confirm the existence of $\ds (2380)$
%and its structure.

Finally, it should also be mentioned that the existence of $\ds$
should further be checked in other experimental processes. Now,
except the $\gamma (or~ e) d$ reaction and $pp$ collision, the
strong decay of the hidden heavy flavor meson, such $b\bar{b}$ meson
and $c\bar{c}$ meson, is also a place to hunt for $\ds$. In
particular, searching for its anti-particle $\bar{\ds}$ in these
processes is even plausible, because the anti-deuteron $\bar{d}$ and
consequently $\bar{\ds}$ can only be created from quark-pair
productions, so that the background would be very clean~\cite{LIHB}.
Now, at $\sqrt{s}\approx 10.6$GeV, the integrated luminosity is
about 470$fb^{-1}$ at BaBar, and about 3$fb^{-1}$ at CLEO. And both
Collaborations have observed $\bar{d}$ production at
$\sqrt{s}\approx 10.6$ GeV~\cite{babar-2014,cleo-2007}. Thus, one
might search for $\bar{d}^*$ in the $\Upsilon(nS) \to \bar{\ds}+p+n$
process. Moreover, the Belle Collaboration has collected even more
data of about 1000 $fb^{-1}$ around $\sqrt{s} \approx 10.6$ GeV, and
they might have the chance to observe the $\bar{d}$ and $\bar{d}^*$
productions in the similar process. Also, BEPCII/BESIII has reached
an integrated luminosity of 1$fb^{-1}$ at $\sqrt{s}=4.42$ GeV and
0.57$fb^{-1}$ at $\sqrt{s}=4.6$ GeV. It might be possible to detect
$\bar{\ds}$ in the $e^+ +e^- \to \bar{\ds}+p+n$ process as well. If
one could observe $\ds$ in the data set accumulated by BaBar, Belle,
CLEO, and BEPCII/BESIII, it would definitely be helpful in
confirming the existence of $\ds$(2380) and its structure.

\begin{acknowledgments}

We would like to thank Stanley Brodsky, Alfone Buchmann, Chao-Hsi
Chang, H. Clement, Haibo Li, Chengping Shen, and Qiang Zhao for
their useful and constructive discussions. This work is partially
supported by the National Natural Sciences Foundations of China
under the grant Nos. 11035006, 11165005, 11475192, 11475181, the
fund provided to the Sino-German CRC 110 ``Symmetries and the
Emergence of Structure in QCD" project by the DFG, and the IHEP
Innovation Fund under the No. Y4545190Y2.
\end{acknowledgments}

\appendix\section{Spin-isospin part}
The spin matrix element in the calculation is \eq I_S&=&\sum
C_{S_Am_A,S_Bm_B}^{S_{AB}m_{AB}}
C_{S'_Am'_A,S'_Bm'_B}^{S'_{AB}m'_{AB}}
C_{S_Am_A,1\mu}^{S'_{A}m'_{A}} C_{S_Bm_B,1\nu}^{S'_{B}m'_{B}}\\
\nonumber &=&\sum (-)^{S_{AB}'-S_{AB}+S_A+S_B-S_A'-S_B'}
\hat{S}'_{A}\hat{S}'_B\hat{S}_{AB}\hat{j}_{23}^2\hat{j} \left \{
\begin{array}{ccc}
1 &S_A &S_A'\\
S_B' &S_{AB}' &j_{23}
\end{array} \right \}
\left \{ \begin{array} {ccc}
1 &S_B &S_B' \\
S_A &j_{23} &S_{AB}
\end{array}
\right \}\\ \nonumber
&\times& \left \{ \begin{array} {ccc}
S_{AB} &1 &j_{23}\\
1      &S_{AB}' &j
\end{array} \right \}
\cdot C_{1\nu,1\mu}^{jm_j}C_{S_{AB}m_{AB}, jm_j}^{S_{AB}'m_{AB}'}\\
\nonumber &=&I_S^0 \cdot C_{1\nu,1\mu}^{jm_j}C_{S_{AB}m_{AB},
jm_j}^{S_{AB}'m_{AB}'},
\en
where $\hat{a}=\sqrt{2a+1}$. For the
present process, the initial $\ds$ and final deuteron have $S_{AB}=3$
and $S_{AB}'=1$, respectively. Moreover, $\Delta$ and nucleon have
$S_A=S_B=3/2$ and $S_A'=S_B'=1/2$, respectively. Therefore, in the
present case $j_{23}=j=2$ is restricted.

Moreover, one can deal with the isospin matrix element similarly.
The isospins of $\Delta$, nucleon, $\ds$, and deuteron are 3/2, 1/2,
0, and 0, respectively. Then $j=0$ and $j_{23}=1$ are constraint
for isospin part.

\end{document}